\documentclass{ws-ijmpc}

\usepackage{epsfig}
\usepackage{latexsym,amsmath}

\newcommand{\av}[1]{\left\langle#1\right\rangle}

\begin{document}

\markboth{Alireza Namazi, Andreas Schadschneider}
{Statistical Properties of online auctions}

\catchline{}{}{}{}{}
%%%%%%%%%%%%%%%%%%%%%%%%%%%%%%%%%%%%%%%%%%%%%%%%%%%%%%%%%%%%%%%%%%%%%

\title{STATISTICAL PROPERTIES OF ONLINE AUCTIONS}

\author{ALIREZA NAMAZI}

\author{ANDREAS SCHADSCHNEIDER}

\address{Institut f\"ur Theoretische Physik, Universit\"at zu K\"oln\\
50937 K\"oln, Germany\\
\{an,as\}@thp.uni-koeln.de}

\maketitle

\begin{history}
\received{\today}
\revised{Day Month Year}
\end{history}

\begin{abstract}
We characterize the statistical properties of a large number of
online auctions run on eBay.
Both stationary and dynamic properties, like distributions
of prices, number of bids etc., as well as relations between these
quantities are studied. The analysis of the data reveals
surprisingly simple distributions and relations, typically of
power-law form.
Based on these findings we introduce a simple method to identify suspicious
auctions that could be influenced by a form of fraud known as shill bidding.
Furthermore the influence of bidding strategies is discussed.
The results indicate that the observed behavior is related to
a mixture of agents using a variety of strategies.

\keywords{Complex systems; emergent behaviour; socio-economic systems;
auctions}
\end{abstract}

%\ccode{PACS Nos.: 89.65.Gh, 89.20.-a, 89.20.Hh, 89.75.-k, 05.45.Tp}

\section{Introduction}

The Internet has triggered new forms of commerce. Not only has
it become possible to order almost every product comfortably from your
home, but also different forms of trading have become popular.
Here especially the success of online auction sites is remarkable.
Through the Internet participation in auctions is no longer
just for a minority, but allows millions of consumers all over
the world to buy and sell almost any kind of product.
Nowadays, auction sites rank very high in the number
of visitors, and also in the average time spent per visit.

Understanding the interactions between users is not only
interesting from an economic point of view.
Surprisingly, despite the success of econophysics\cite{stanley,econobook},
only very few studies of the statistical properties of
online auctions by physicists exist\cite{IY,bornholdt}.
Ref.~\refcite{IY}  focused on properties of agents (i.e.\ bidders and sellers)
in auctions, e.g. the number of agents participating or
the frequency of bids placed. It was found that online auctions
can be considered as complex systems that exhibit emergent behaviour.
In Ref.~\refcite{bornholdt}, the relation between
different agents was investigated by analyzing the structure
of the interaction network resulting from online auctions.

Online auctions are conducted in a different way than the standard
English auction which usually ends when nobody is willing to bid
higher ("soft close"). In contrast, online auctions end at a fixed
time known to all agents ("hard close").  Any agent offering an item
at an online auction house like eBay first has to specify the
starting time, starting price and duration of the auction. Each new
bid placed by another agent has to exceed the currently listed price
by a preset increment. Agents can either bid incrementaly by placing
a bid that corresponds to the current price plus the minimal
acceptable increment, or take advantage of {\em proxy bidding}. In
proxy bidding an agent indicates the maximum price he/she is willing
to pay for the given item (proxy price), which is not disclosed to
other bidders. Each time a bidder places a new bid, the auction
house makes automatic bids for the agent with an active proxy bid
\footnote{Note that at most one {\em active} proxy bid can exist at
any time.}, outbidding the last bid with a fixed increment until the
proxy price is reached. The agent with the highest bid wins.

Due to the proxy system eBay auctions are very similar to {\em second
  price auctions}\cite{vickrey}. The winner (buyer) does not pay a
price corresponding to his actual winning bid. Instead the final price
is determined by the second highest bid, plus the preset increment.

From a physics point of view the development of a simple microscopic model
for online auctions appears to be quite challenging. In contrast
to most other processes it is essential that the dynamics ends at
a certain time and that this fact is known to all the agents. As we
will show below, this is clearly reflected in the empirical data.
We will not deal with the problem of modelling here, but instead try to
determine the generic statistical properties of online auctions
empirically. These data might then be used in order to test the
validity of model approaches.

eBay keeps a detailed record of the bidding history
that is publicly available.
Bidding agents are distinguished by a unique user name. This allows
to study the dynamics of the bidding process in a quantitative way.
In this paper, we analyze and characterize the statistical properties
of online auction data. We focus on the properties of the auctions
and especially the relations between various quantities (price, number
of bids placed etc.). Furthermore we also study the dynamics of
the bidding process, not only the stationary properties after the end
of the auction. This allows interesting insights into the strategies
used by the agents participating in the auction.

\section{Data collection} 

For our investigation we have used two major sets of auction data from
eBay Germany ({\tt www.ebay.de}) that allow us to focus on various
properties in more detail.
Many of our results are found to be valid for both
data sets.

The first set (DB-1) comprises data collected from auctions running on
March 22, 2004. We focus on auctions with the label
"OVP" \footnote{Abbreviation for 'Originalverpackung', meaning
'original wrapping' or 'sealed'.} in the title, indicating a new
product. The data comprises 173,315 auctioned items, grouped in 9904
subcategories by eBay. 262,508 distinct agents bidding on items and
43,500 sellers offering auctioned items are identified.

The second set (DB-2) are data collected over 10 months in the
subcategories "web projects" and "websites \& domains". The auctions
involve 11,145 agents bidding on 52,373 items.

The data set DB-1 allows to study 
auctions where the majority of products has a well-defined value
(e.g.\ market price) for all agents.  In contrast, the set DB-2
comprises items where the value can be different for different agents.

\section{Statistical analysis}
Using these data one finds that the distribution of the total number
of bids placed by a given agent follows  a power-law for both data sets.
The total number of distinct items offered by a given agent (as seller)
also follows a power-law distribution.

However, first we have examined distributions of static properties
(i.e.\ properties at the end of an auction). A good qualitative
agreement with the results of Ref.~\refcite{IY} is observed. E.g. the
distributions of the distinct number of agents $n_{\rm agent}$
simultaneously bidding on a certain item and the total number of bids
$n_{\rm bids}$ received for an item both follow exponential
distributions $P(n)\sim\exp(-n/n_{0})$, where $n_0=6.5$ for
$n_{\rm bids}$ and $n_0=2.9$ for $n_{\rm agent}$.
This is in agreement with Ref.~\refcite{IY}, where the values
$n_0=5.6$ for $n_{\rm bids}$
and $n_0=2.5$ for $n_{\rm agent}$ for eBay and $n_0=10.8$ for
$n_{\rm bids}$ and $n_0=7.4$ for $n_{\rm agent}$ for eBay Korea were
obtained.

The activity of individual agents as bidder or seller
follows power-law distributions\cite{drarbeit}.
One can find that the distribution of the total number of bids placed by
the same agent, denoted by $n_{\rm bids}$, follows a power law
\begin{equation}
P(n_{\rm bids})\sim n_{\rm bids}^{-\gamma},
\end{equation}
where $\gamma=1.9$.

\begin{figure}
\centerline{\epsfxsize=0.95\columnwidth \epsfbox{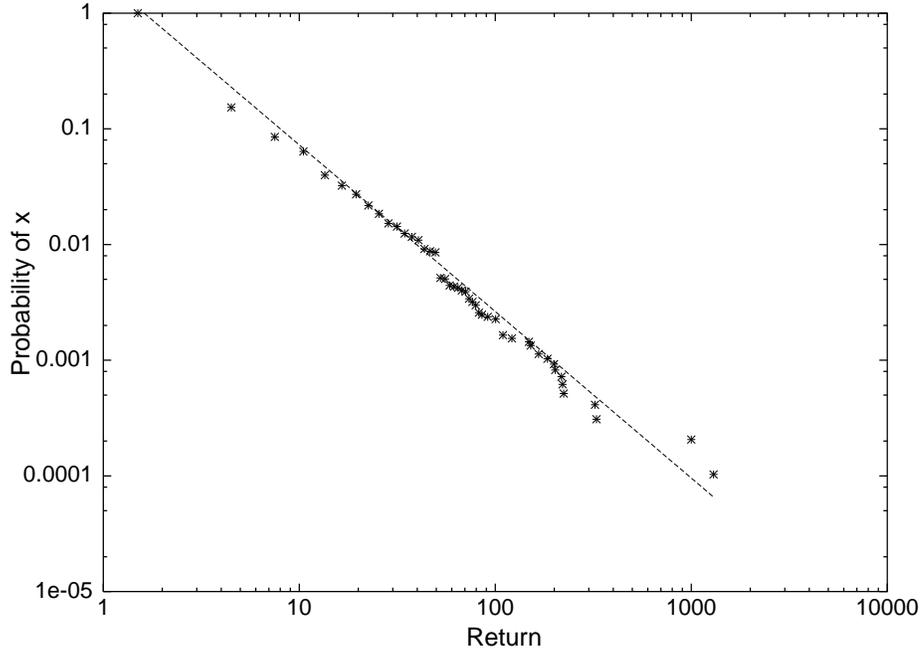}} 
\caption{The cumulative probability distribution of returns $\varrho$
follows a power law with exponent $-1.44$. } 
\label{FigReturnDis}
\end{figure}

In order to quantify the bidding process we define a {\em dimensionless}
variable that we call return $\varrho$. It is the
relative increase of the submitted bid $b$:
\begin{equation}
\varrho =  \frac{b- p_{\rm current}}{p_{\rm current}}, \label{eqreturn}
\end{equation}
where $p_{\rm current}$ is the current or listed price just before 
the bid is placed. In analogy to the quantity used in 
financial markets\cite{stanley,econobook} it measures relative changes.
Very large values of the return are usually generated by the
first submitted bid, if the bidder follows the recommendations of 
eBay and submits the maximum price (s)he is willing to pay.
The distribution of $\varrho$ is found  to follow a power law for
almost three orders of magnitude with exponent $-2.44$
(Fig.~\ref{FigReturnDis}).
The form of the distribution appears to be quite stable even for
different time spans of the bid submission\cite{drarbeit}.

Although the bids are correlated, the returns show only very short-ranged
correlations. By computing the correlation function  $c_{ij}=
\av{\varrho_{i}\varrho_{j}}- \av{\varrho_{i}}\av{\varrho_{j}}$
of the returns (indices $i$ denote the
chronological order of arriving bids; averaging is done over all
auctions) one finds that off-diagonal terms $c_{ij}$ 
with $i\neq j$ are very small\cite{drarbeit}. This result
is rather surprising and we have currently no simple explanation.

One of the interesting questions is the functional dependence of the price
on other dynamic parameters such as number of bids. Several
studies\cite{Melnik,Lucking,Jank} investigate the influence
of static parameters like the ending time (which day of the week,
on which daytime), start price, reputation of the seller etc.
Much less is known (quantitatively) about relations between
dynamic parameters.

\begin{figure}
\centerline{\epsfxsize=0.95\columnwidth  \epsfbox{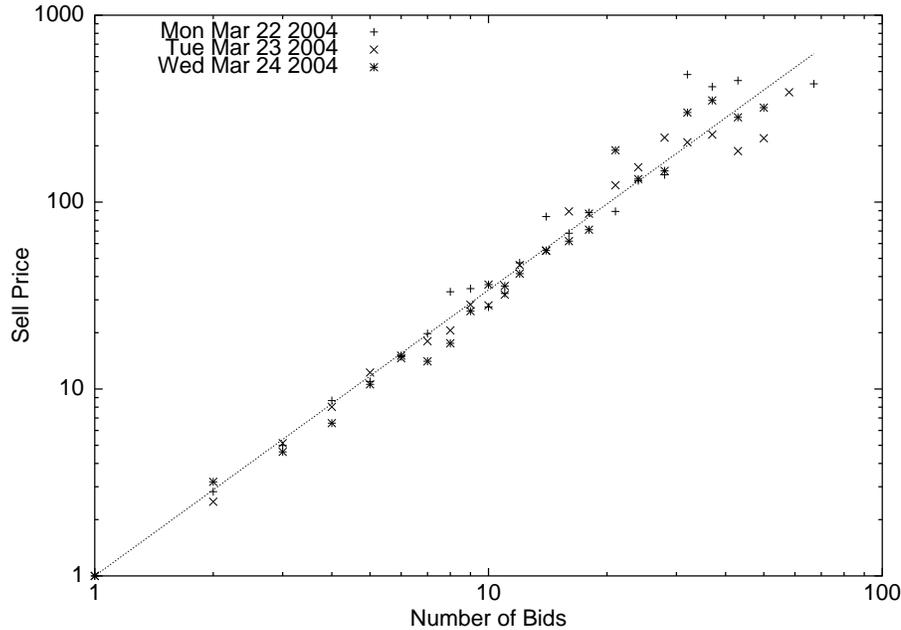}} 
\caption{Relation between final price $\av{p_{\rm final}}$ (in Euro, 
    for starting price 1~Euro) 
    and number of bids $n_{\rm bid}$ for different days based on subsets 
    of DB-1. 
  The straight line is a power-law fit with exponent $\alpha=1.53$. }
\label{FigPB6D}
\end{figure}

We have determined the dependence of the price on other dynamic
parameters. For the relation between the final price $p_{\rm final}$
and number of bids $n_{\rm bid}$ placed on the item a power law
is found:
\begin{equation}
\av{p_{\rm final}} \propto n_{\rm bid} ^{\alpha}, \label{PB}
\end{equation}
where $\alpha=1.58$ for the data set DB-2 and $\alpha=1.53$ for the
data set DB-1 (Fig.~\ref{FigPB6D}).

This functional dependence seems to be universal and independent of the
subcategories or time intervals used for data collection\cite{drarbeit}.
Especially there seems to be no significant difference between
the two data sets DB-1 and DB-2.

Empirically we find that the distribution of the final price for
fixed total number of bids is well described by a log-normal
distribution\cite{drarbeit}. A possible explanation uses
the observation that the returns $\rho_i$ are almost uncorrelated.
Assuming $b=P_n$ in (\ref{eqreturn}), the price $P_n$ after the
$n$-th bid is given by a multiplicative process of the form
$P_n = (\varrho_{n}+1)P_{n-1}$. Since the returns $\varrho_{n}$
are found to be uncorrelated,
$\ln(P_n) = \sum_{i=1}^n \ln(\varrho_{i}+1)$ 
converges to a Gaussian distribution for large $n$.

In the following we discuss a possible origin of systematic deviations
from the average behaviour described above, namely fraud.

\subsection{Shill bidding}

It is not uncommon that certain agents try to manipulate the sell
price of the items they offer. Therefore we distinguish two different
types of bidding behaviour: 1) All bidders try to keep the price as
low as possible; 2) At least one bidder tries to push the price
higher.  The second type of behaviour is known as shill bidding. Shill
bidding (also known as "bid padding") is the deliberate use of
secondary registrations, aliases, family members, friends, etc.
to artificially drive up the price of an item.  This is
strictly forbidden by eBay\cite{EB}, but nevertheless happens quite
frequently. Usually this sort of manipulation can be identified only
after the auction has ended because the whole purpose of shill bidding
is increasing the price without winning in the end! 
Shill bidding occurs mostly in 
auctions of products that do not have a well-defined value,
e.g.\ market price.

\begin{figure}
\centerline{\epsfxsize=0.99\columnwidth \epsfbox{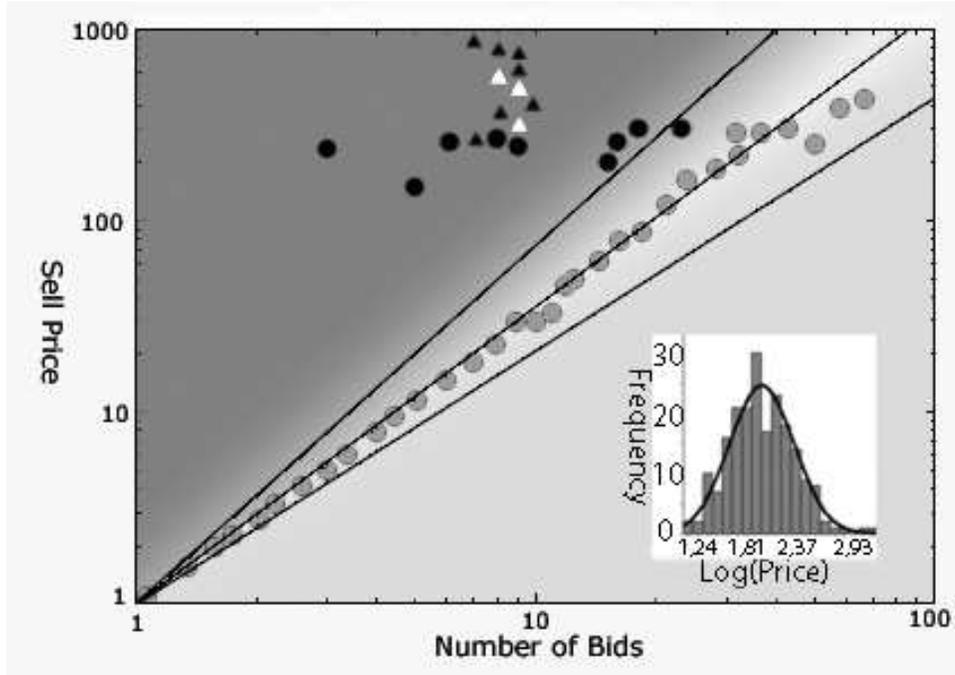}}
\caption{Distribution of sell prices as function of the number
  of bids placed.  Only auctions with a starting price of 1 Euro are
  considered. Gray dots correspond to the average sell price for fixed
  number of bids. The broken lines specify the confidence interval
  (one standard deviation) determined from a log-normal fit (see
  inset). This is shown for $n_{\rm bid}=20$ in the inset. For other
  values of $n_{\rm bid}$ very similar results are obtained. 
  Black dots indicate auctions identified as shill bidding 
  using the criteria of Refs.~12-14 (test a). 
  Triangles denote auctions that have been tested for possible shill 
  bidding (test b). For black triangles strong indications for shill 
  bidding have been found,  whereas auctions corresponding to white 
  triangles are unsuspicious. Light-gray and
  dark-gray colours denote regimes with high or low probability of
  shill bidding, respectively.}
\label{fig6}
\end{figure}

We have tried to identify this sort of manipulation through the
statistical properties. Indeed, successful shill bidding leads to
auctions which show clear deviations from the observed simple
statistical laws. To quantify this we have defined a confidence
interval based on the assumption that the price follows a log-normal
distribution (see inset in Fig.~\ref{fig6}). The variance $\sigma
^2$ of the Gaussian distribution of the logarithm of prices with a
fixed number of bids characterizes the confidence interval which is
indicated in Fig.~\ref{fig6} by the two straight lines with slopes
$1.32$ and $1.92$, respectively.

We have performed two tests
by a) investigating the statistical properties of auctions identified
e.g.\ from discussion forums \cite{EB5}
as shill bidding, and b) by checking whether randomly chosen auctions
outside the confidence interval of the price-bid relationship show
indications of shill bidding. Both tests require an extensive amount
of work, e.g. the investigation of the trading history of the seller
over a long time or monitoring eBay discussion forums\cite{EB5}.
Fig.~\ref{fig6} shows a comparison between these shill-auctions and
the average behaviour of all auctions (DB-1 and DB-2), 
clearly indicating systematic deviations.

For test a) we have chosen 9 auctions that clearly have been
identified as manipulated by shill bidding, e.g. through information
from discussion forums\cite{EB5}. Only after that we have
investigated the bidding history of these auctions in more detail.
Fig.~\ref{fig6} shows that all of those, except for one, are clearly
outside the confidence interval.  For test b) 10 auctions outside the
confidence interval have been chosen randomly. These have been checked
thoroughly for indications of shill bidding. This also required
investigating other auctions by the same seller etc. In this way we
have found a clear indication for shill bidding in 7 of the 10
auctions.

The results of the tests indicate that it would be sufficient to
check the suspicious auctions (i.e.\ those outside of the confidence
interval) in more detail. This would reduce
the complexity of detecting manipulations drastically.
Furthermore these deviations usually can be observed even {\em before}
the end of the auction. Monitoring the current price as function
of the number of submitted bids up to the time of observation,
one can check whether this price is inside or outside of the
confidence interval in Fig.~\ref{fig6}. In the latter case, a
high current price is an indication that bidders should be cautious
(although it is no proof of shill bidding).

\subsection{Bid submission times} 

\begin{figure}
\centerline{\epsfxsize=0.99\columnwidth \epsfbox{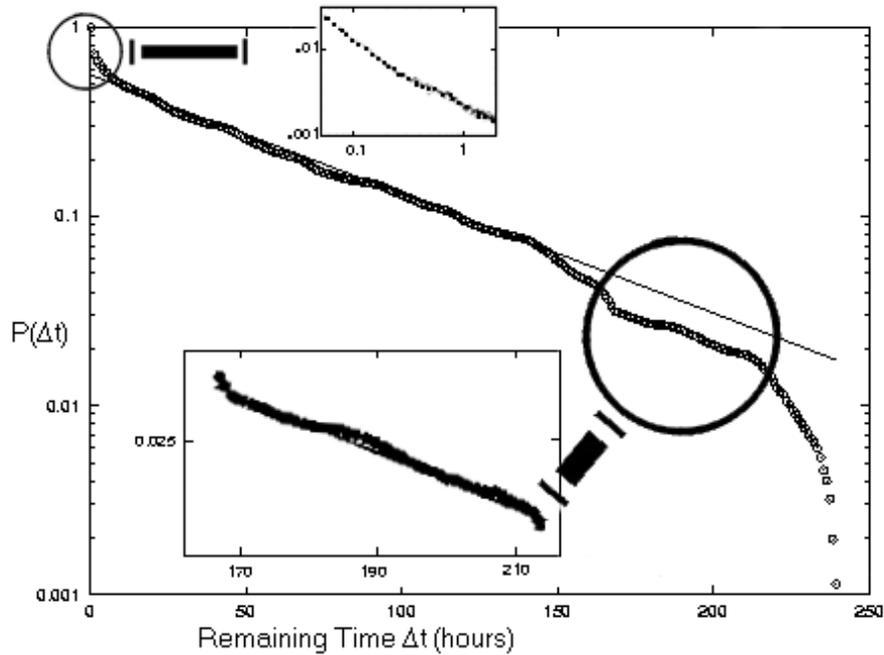} } 
\caption{Cumulative
  distribution of bid submission times as a function of the time
  remaining until the end of the auction (for DB-2).  Two
  regimes with exponential behaviour can be observed related to the
  most common auction lengths of 7 and 10 days (i.e.\ 168 and 240
  hours). Both parts are well described by $P(\Delta t)\sim
  \exp(-\Delta t/T_{0})$ with $T_0=68.94$~h.  Close to the end of the
  auctions sniping leads to an algebraic distribution $P(\Delta t)\sim
  (\Delta t)^{-\gamma}$ with $\gamma =1.1$.  The data set DB-1 shows a
  very similar behaviour with $T_0=90.09$~h and $\gamma =1.15$
  (Ref.~6).  \label{fig10}} 
\end{figure}

Second price auctions were
originally introduced by Vickrey\cite{vickrey} to make everyones
bidding strategy independent of strategies used by the other
bidders.  It could be shown that late bidding or bidding with a
price less than one is willing to pay are not optimal strategies.
Using game theory based models\cite{RO02,RO05,wilcox,BB04}, it
seems hard to understand why bidders bid more than once. We analysed
the collected data and found that bidders prefer to bid close to
auction ending times.  Fig.~\ref{fig10} shows the cumulative
distribution of bid submission times as a function of the time
remaining until the end of the auction. Two regimes with exponential
behaviour can be observed related to the most common auction lengths
of 7 and 10 days. Both parts are well described by $P(\Delta t)\sim
\exp(-\Delta t/T_{0})$ with $T_0=68.94$~h.  Close to the end of the
auctions so-called {\em sniping} leads to an algebraic distribution
$P(\Delta t)\sim (\Delta t)^{-\gamma}$ with $\gamma =1.1$.  
Sniping is a special bidding strategy where the agent tries to submit a
winning bid just before an auction closes in order to prevent other
bidders from outbidding the sniper or driving the price higher. In fact, 
game-theoretic arguments show that sniping can be advantageous\cite{BB04}, 
e.g.\ to avoid bid wars or auction fraud due to shill bidding.

Surprisingly, the distribution of bid submissions times
is qualitatively very similar both data sets DB-1 and DB-2. 
Although the scales in the exponential part differ slightly
($T_0=68.94$~h for DB-2 and $T_0=90.09$~h for DB-1), the
exponents of the algebraic part are almost identical.
From earlier investigations\cite{RO02} one could have expected
a qualitative difference between the two data sets.

\section{Discussion} 

eBay online auctions can be considered as
strongly correlated processes.  Correlations are due to influences of
standard market prices, supply and demand mechanisms within eBay
auctions and seller's feedback information available for any visitor
of the site\cite{FEEDBACK}.  We have shown several emerging
properties, which could be seen as an evidence of complexity.  The
generic statistic behavior of agents leads to a very simple relation
between price and number of bids.  As an application we have found
that a kind of fraud known as shill bidding leads to significant
deviations from the average behaviour.

Economic theories have mainly used game theoretic methods to describe
auctions\cite{RO02,RO05,wilcox,BB04}.
These are based on the assumption of {\em rational} agents
using specific bidding strategies,
like evaluation (placing one's true value, usually early in the auction), 
unmasking (bidding as long as someone else is the highest bidder, e.g.\
to determine the highest bid),
or incremental bidding (placing minimum acceptable bids).
Our results indicate
that eBay can be regarded as a complex system consisting of
a mixture of agents using different strategies. 
In future work we intend to determine the influence of the rationality
of the agents (in the sense of game theory).

As mentioned in the introduction, to our knowledge so far
no microscopic stochastic model for the description of
online auctions exists. Most approaches use game theoretical methods
which focus on the effects of different strategies.
Our empirical data indicate that a model needs to take into account 
that a mixture of agents with different strategies exists which
is responsible for the observed behaviour. This can be seen most
clearly in the distribution of bid submission times. 
In order to take into account the hard close of online auctions
a naive approach would use dynamical rules which are itself time-dependent,
which would be a challenge for theoretical studies of the model.

\section*{Acknowledgments}
We thank Axel Ockenfels and Maya Paczuski for helpful discussions.

\end{document}